\documentclass[twocolumn,pre,floatfix]{revtex4}
\usepackage{psfrag,epsfig,amsfonts,amssymb,amsmath,wasysym,bm}
\usepackage{dcolumn}
\usepackage{bbold}
\usepackage[normalem]{ulem}
\usepackage{color}
\usepackage{tabularx}
\usepackage{tikz}

\usepackage[inline]{enumitem} 

\usepackage{hyperref}

\newcommand{\ket}[1]{\lvert #1 \rangle} 	
\newcommand{\bra}[1]{\langle #1 \rvert}	
\newcommand{\ketL}[1]{\lvert #1 \rangle_{\! \lambda}} 	
\newcommand{\braL}[1]{{_\lambda\!}\langle #1 \rvert}	

\newcommand{\<}{\left\langle}	
\renewcommand{\>}{\right\rangle}	

\newcommand{\e}{\mathrm{e}}		
\newcommand{\I}{\mathrm{i}}		

\newcommand{\lvsp}{\varepsilon}

\newcommand{\Ath}{\langle A\rangle_{\!\!\;\rm{mc}}}
\newcommand{\varV}{\sigma^2}
\newcommand{\cmagV}{\bar\sigma}
\newcommand{\cvarV}{\cmagV^2}
\newcommand{\bwV}{\Delta_v}

\newcommand{\rhoMC}{\rho_{\mathrm{mc}}}

\newcommand{\RR}{{\mathbb R}}
\newcommand{\CC}{{\mathbb C}}

\newcommand{\tr}{\mbox{Tr}}

\providecommand{\av}[1]{[#1]_V}
\providecommand{\avv}[1]{\left[#1\right]_{\!V}}

\renewcommand{\d}{\mathrm{d}} 
\newcommand{\lmat}{\left( \begin{matrix}}	
\newcommand{\rmat}{\end{matrix} \right)}	

\newcommand{\sigmaName}{perturbation profile}
\newcommand{\uName}{overlap distribution}
\newcommand{\gName}{response profile}

\definecolor{LDcolor}{RGB}{0,196,32}

\definecolor{revcolor}{RGB}{64,96,255}

\begin{document}
\title{Modification of quantum many-body relaxation \\ by 
perturbations exhibiting a banded matrix structure}
\author{Lennart Dabelow}
\author{Patrick Vorndamme}
\author{Peter Reimann}
\affiliation{Fakult\"at f\"ur Physik, 
Universit\"at Bielefeld, 
33615 Bielefeld, Germany}
\date{\today}

\begin{abstract}
We investigate how the observable relaxation behavior of an isolated quantum many-body system
is modified in response to weak-to-moderate perturbations within a nonperturbative typicality framework.
A key role is played by the
so-called \sigmaName,
which characterizes the dependence of the
perturbation matrix elements in the
eigenbasis of the unperturbed Hamiltonian
on the difference of the 
corresponding energy eigenvalues.
In particular, a banded matrix structure is quantitatively captured by
a \sigmaName\ 
which approaches zero for large energy differences.
The temporal modification of the relaxation is linked to the \sigmaName\ 
via a nonlinear integral equation, which admits approximate analytical
solutions for sufficiently weak and strong perturbations, and for which
we work out a numerical solution scheme in the general case.
As an example, we consider a spin lattice model 
with a pronounced banded matrix structure,
and we find very good agreement of the numerics with
our analytical predictions without any free fit parameter.
\end{abstract}

\maketitle

\section{Introduction}
\label{sec:Intro}

Despite their microscopic chaoticity \cite{haa10,dal16} the macroscopically observable behavior of isolated quantum many-body systems is often surprisingly regular.
For instance, it is by now well-established that these systems generically equilibrate and usually even thermalize \cite{gog16, dal16, mor18}, and that the approach to equilibrium 
quite often
 follows a rather simple and direct route.
Understanding how this dynamics emerges from a microscopic description, however, is still a theoretical challenge attracting considerable attention recently.
A particularly interesting question in this context is how the observable relaxation behavior of a given system is modified under the influence of reasonably weak perturbations, linking, for example, analytically tractable simple systems (e.g.\ noninteracting, integrable) to generic ones (e.g.\ interacting, nonintegrable).

Characterizing the response of a given system to a perturbation is a recurrent problem in many areas of physics.
Arguably the standard approach is to expand the pertinent equations of motion in terms of the perturbation strength and to solve the resulting hierarchy of simplified equations iteratively.
Unfortunately, such a strategy is doomed to failure in the case of quantum systems with many degrees of freedom.
Due to their extremely dense energy spectra, the concomitant small denominators of a perturbative expansion limit its applicability to extremely short time scales much smaller the observed relaxation times.
While there is strong evidence that related concepts like Fermi's golden rule and linear response theory can describe many-body dynamics in certain scenarios \cite{bac18, mal19thermal, ric19}, this somewhat surprisingly holds \emph{despite} the many-body character and not because of it.

Here we tackle the question of how a many-body systems responds to perturbations by ``nonperturbative'' methods, namely a typicality approach that aims to extract and separate the macroscopically relevant perturbation characteristics from the huge number of microscopic degrees of freedom.
Our starting point is an isolated many-body quantum system described by a time-independent 
{\em reference Hamiltonian} $H$, and prepared in some initial state far from equilibrium.
Provided that we know the observable relaxation dynamics of this unperturbed reference system, we ask how the behavior is changed when adding a weak-to-moderate perturbation $\lambda V$.
In other words, the system still starts from the same initial state, but now evolves in time according to the perturbed Hamiltonian
\begin{equation}
\label{eq:H}
	H_\lambda := H + \lambda V \,.
\end{equation}
One situation that could be modeled by such an approach is an unperturbed system composed of two isolated subsystems at equilibrium, which are then coupled sufficiently weakly via the perturbation $V$ and relax to a new, joint equilibrium state.
Another interesting scenario arises when the reference system $H$ is integrable, in which case one can often calculate the unperturbed behavior analytically.
In particular, integrable systems usually still equilibrate (just like the nonintegrable ones),
meaning that expectation values of 
experimentally relevant
observables approach a constant value and stay there for most of all later times.
However, these integrable systems (unlike the nonintegrable ones) may not thermalize, i.e., equilibrium expectation values are not described by the pertinent thermodynamic equilibrium ensemble
and call for extensions like generalized Gibbs ensembles instead \cite{rig07, kol11, vid16, ess16}.
Adding a small integrability-breaking perturbation commonly leads to ``prethermalization'' 
\cite{ber04, moe08, lan16, mor18, rei19pretherm, mal19thermal}, 
meaning that the system still follows the unperturbed (nonthermalizing)  behavior for 
quite some time before eventually departing towards the associated thermal state.
As a third example, more generally, one may think of the unperturbed system as some system for which the relaxation dynamics happens to be known, and ask for the behavior when changing some parameter of the Hamiltonian (e.g.\ a ``quantum quench'' \cite{gog16, ess16, mit18}).

Basing our analysis on previous results from Ref.~\cite{dab20relax}, we recap those findings in Secs.~\ref{sec:Scope} and~\ref{sec:TypResponse}.
More precisely, we introduce the considered classes of systems and formulate the key assumptions of our theory in Sec.~\ref{sec:Scope},
and establish the announced theoretical prediction of the many-body response in Sec.~\ref{sec:TypResponse}.
A crucial role is played by the resolvent $(z-H_\lambda)^{-1}$ averaged over an ensemble of perturbed Hamiltonians $H_\lambda$, whose computation we expound in Sec.~\ref{sec:Resolvents}.
These results are then used in Sec.~\ref{sec:Examples} to make the prediction from Sec.~\ref{sec:TypResponse} explicit and to compare it to numerical examples for random-matrix and spin models.
Finally, we summarize and discuss our results in Sec.~\ref{sec:Conclusions}.

\section{Scope and prerequisites}
\label{sec:Scope}

Before presenting our main result, we introduce the setting and collect several 
key assumptions about the physical situations we aim to describe
(see also Supplemental Material of Ref.~\cite{dab20relax} for further technical details).

The isolated many-body quantum system of reference is described by a 
time-independent Hamiltonian $H = \sum_\nu E_\nu \, \ket{\nu} \bra{\nu}$ and
is prepared in some (pure or mixed, and generally far from equilibrium) 
initial state with density operator $\rho(0)$.
According to textbook quantum mechanics, the state at any later time is then given by $\rho(t) = \e^{-\I H t} \rho(0) \e^{\I H t}$ ($\hbar = 1$).
Of primary interest to us are the time-dependent expectation values $\< A \>_{\!\rho(t)} := \tr[ \rho(t) A ]$
of self-adjoint operators $A$ which model some experimentally or theoretically 
relevant observable, such as (sums of) local and/or few-body operators 
\cite{dal16,gog16,mor18}.
Similarly, the time-evolved state of the perturbed system with Hamiltonian $H_\lambda$ from~\eqref{eq:H} is given by $\rho_\lambda(t) = \e^{-\I H_\lambda t} \rho(0) \e^{\I H_\lambda t}$.

Overall, the main objective of our present work is to establish quantitative predictions for
the perturbed dynamics $\langle A \rangle_{\!\rho_\lambda(t)}$ based on the unperturbed behavior $\langle A \rangle_{\!\rho(t)}$ and some essential characteristics of the perturbation $V$.

Regarding the systems under study, the following 
four key assumptions will be taken for granted hereafter:

\begin{enumerate*}[label={(\roman*)}, mode=unboxed, itemjoin=\\ \indent]
\item \label{lst:Assumptions:EnergyWindow}
The system should exhibit a well-defined macroscopic energy,
implying that the initial state $\rho(0)$ (and hence also $\rho(t)$ at any later $t$) only significantly populates levels $E_\nu \in I$ within a macroscopically small 
{\em energy window} $I := [\mathcal{E}, \mathcal{E} + \Delta]$.
In particular, it is assumed that the density of states (DOS)
\begin{equation}
\label{eq:DOS}
	D(E) := \sum_{\nu : E_\nu \in I} \delta( E - E_\nu )
\end{equation}
is approximately constant throughout $I$, $D(E) \approx \varepsilon^{-1}$ with the 
{\em mean level spacing} $\varepsilon$.
At the same time, the system's many-body character entails that the window $I$ is still microscopically large in the sense that the number of levels contained in $I$ is exponentially large in the system's degrees of freedom \cite{lan70}.
We emphasize that the initial state does \emph{not} necessarily \emph{define} the window $I$.
On the contrary, the window $I$ can be a to some extent arbitrary interval with the two prerequisites that it exhibits an approximately constant DOS and contains all $E_\nu$ with nonnegligible 
level populations $\langle \nu |\rho(0)|\nu\rangle$.

\item \label{lst:Assumptions:WeakPerturbation}
The perturbation should be sufficiently weak so as to leave the thermodynamic properties of the system basically unchanged.
Notably, phase transitions induced by the perturbation are thus ruled out.
Recalling from textbook statistical mechanics that the DOS in~\eqref{eq:DOS} is related to 
the Boltzmann entropy $S(E)$ via $D(E) = \e^{S(E)/k_{\mathrm B}} S'(E)$ with Boltzmann's constant $k_{\mathrm B}$, this assumption particularly implies that also the DOS of the perturbed $H_\lambda$ 
remains approximately constant with mean level spacing $\varepsilon$ [cf.\ assumption~\ref{lst:Assumptions:EnergyWindow}].
Due to the 
generic level repulsion of
interacting many-body systems \cite{haa10}, the spectrum of $H_\lambda$ is typically indeed rather stiff, 
meaning that the individual eigenvalues exhibit very fast fluctuations upon variation of $\lambda$, 
while their density 
only changes very slowly \cite{note:GroundState}.

\end{enumerate*}

We remark that such negligible changes of the thermodynamic properties do
not rule out interesting and nontrivial changes of the
relaxation dynamics, notably if the unperturbed 
Hamiltonian
is in some sense special (e.g., integrable, commuting with $A$ or $\rho(0)$, ...),
see also the examples below Eq.~\eqref{eq:H} and in Sec. \ref{sec:Examples}.

\begin{enumerate*}[resume, label={(\roman*)}, mode=unboxed, itemjoin=\\ \indent]
\item \label{lst:Assumptions:StrongPerturbation}
The perturbation should be sufficiently strong so that it significantly mixes a large number of unperturbed levels.
Denoting by 
$\ketL{m}$ the eigenvectors of
$H_\lambda$ in (\ref{eq:H}), this is to say that the overlaps
\begin{equation}
\label{eq:U}
	U_{m \nu} := \braL{m} \nu \rangle
\end{equation}
between the unperturbed and perturbed eigenvectors should extend across
a scale $\Gamma$ with $\Gamma \gg \varepsilon$, i.e., $U_{m\nu}$ should 
be nonnegligible (in a coarse-grained sense, see below) as long as 
$\lvert E_m - E_\nu \rvert \lesssim \Gamma$
(see also Eq. (\ref{eq:u})).
On the other hand, note that assumptions~\ref{lst:Assumptions:EnergyWindow} and \ref{lst:Assumptions:WeakPerturbation} practically require
$\Gamma \ll \Delta$
\cite{note:StateBroadening},
where $\Delta$ is the width of the energy window $I$ from \eqref{eq:DOS}.
The extreme density of levels of typical many-body systems [see below Eq.~\eqref{eq:DOS}] still leaves room for a large range of parameters $\lambda$ such that $\varepsilon \ll \Gamma \ll \Delta$.
In particular, we can and will take for granted that the number of levels $N_v := \Gamma / \varepsilon$ that get mixed by the perturbation is still exponentially large in the system's degrees of freedom $f$ \cite{dab20relax},
i.e.,
\begin{equation}
\label{eq:NumMixedLevels}
	N_v := \Gamma / \varepsilon =  10^{\mathcal{O}(f)} \,.
\end{equation}
\end{enumerate*}
Without going into the details, we remark that perturbations which do {\em not}
satisfy the requirement $\varepsilon \ll \Gamma$ turn out (as one might have expected)
to actually be so weak that they do not notably modify the unperturbed relaxation 
on any reasonable time scale. Incidentally, the same behavior will also
be correctly reproduced by our final results. In this sense, the requirement
$\varepsilon \ll \Gamma$ is not really indispensable.

So far, these considerations have been very general and did not exploit any 
more specific properties of the 
actual system at hand.
To make any progress, it is clear that some information about the perturbation $V$ and possibly also the observable $A$ and initial state $\rho(0)$ must be taken into account.
The common lore of statistical physics furthermore
suggests that, despite its microscopic complexity, the observable behavior of a many-body system 
can usually be described in terms of a 
relatively small number of macroscopic (coarse-grained) quantities,
for instance some appropriately defined (local) densities.
This brings us to our main assumption about the structure 
of admissible perturbations:

\begin{enumerate*}[resume, label={(\roman*)}, mode=unboxed, itemjoin=\\ \indent]
\item \label{lst:Assumptions:Bandedness}
On a coarse-grained level, the magnitude of the perturbation matrix elements $V_{\mu\nu} := \bra{\mu} V \ket{\nu}$ within the energy window $I$ should only depend on the energy difference $\lvert E_\mu - E_\nu \rvert$ of the coupled levels \cite{f1}, i.e.,
\begin{equation}
\label{eq:Bandedness}
	\left[ \lvert V_{\mu\nu} \rvert^2 \right]_{\mathrm{loc}} \simeq \varV(\lvert E_\mu - E_\nu \rvert) \,,
\end{equation}
where $[ \,\cdots\, ]_{\mathrm{loc}}$ denotes a local average over matrix 
elements corresponding to levels that are close to $E_\mu$ and $E_\nu$ 
in energy (see also Sec.~\ref{sec:Examples:Spin} for an explicit example).
Put differently, the left-hand side in (\ref{eq:Bandedness})  
is understood (and formally defined) analogously as when going over,
e.g., from classical point particles to (local) particle densities, namely 
as the effective density of the perturbation's squared matrix elements
(in modulus, and ``local''  with respect to the spectrum of $H_0$).
Accordingly, $\varV(E)$ in (\ref{eq:Bandedness}) is 
denoted as the {\em \sigmaName}, and is,
by construction, a smooth \cite{note:diagV} and slowly 
varying function of $E$ (compared to the mean level 
spacing $\varepsilon$).
\end{enumerate*}

Semiclassical arguments \cite{fei89, fyo96} as well as numerical evidence 
\cite{gen12, beu15, kon15, bor16, jan19} suggest that a rather common feature 
of realistic perturbations is a so-called banded structure of the perturbation matrix 
$V_{\mu\nu}$ (see also Fig.~\ref{fig:SpinModel} in Sec.~\ref{sec:Examples:Spin} 
below for a particular example). By definition, this means that the (coarse-grained) $V_{\mu\nu}$ indeed depend only on $E_\mu - E_\nu$ and that the 
{\sigmaName} $\varV(E)$ in (\ref{eq:Bandedness}) approaches 
zero for $E\to\infty$.
However, it should be emphasized that $\varV(E)$ is also admitted to 
remain finite for $E\to\infty$, i.e., the matrix $V_{\mu\nu}$ may but need 
not exhibit a banded structure \cite{f2}.
Yet another common feature of many realistic perturbations 
is a so-called sparse matrix structure (large fraction of vanishing 
matrix elements $V_{\mu\nu}$),
prominently arising, e.g., if the reference Hamiltonian $H$ is noninteracting 
and $V$ describes few-body interactions \cite{bro81, fla97, bor16, fyo96}.
Again, our present approach 
is still compatible with a possibly (but not necessarily) 
sparse structure of $V_{\mu\nu}$ [the local average
in (\ref{eq:Bandedness}) then must extend over many 
nonvanishing matrix elements].

Our next goal is to establish the key role of the \sigmaName~\eqref{eq:Bandedness}
for the deviations of the perturbed expectation values $\langle A \rangle_{\!\rho_\lambda(t)}$ 
from the unperturbed $\langle A \rangle_{\!\rho(t)}$.
The main idea is to consider not one particular $V$, but rather
 an entire ensemble of perturbations, all of which share the property~\eqref{eq:Bandedness} with the ``true'' perturbation of interest, but are otherwise unbiased and rather arbitrary.
More precisely, apart from the trivial constraint $V^*_{\mu\nu} = V_{\nu\mu}$, we choose the matrix elements $V_{\mu\nu}$ to be independent random variables following a probability distribution
\begin{equation}
\label{eq:VProb}
	p_{\mu\nu}(v) := \av{ \delta(V_{\mu\nu} - v) } = f_{\vert E_\mu - E_\nu \rvert}(v) \,,
\end{equation}
where $\av{\,\cdots}$ denotes the average over the ensemble of perturbations, and $\{ f_E(v) \}_{E>0}$ is a family of probability densities on $\mathbb R$ or $\mathbb C$ with mean zero and variance $\varV(E)$.
Likewise, $f_0(v)$ is a probability density on $\mathbb R$ of vanishing mean and finite variance \cite{note:diagV}.
Note that the ensemble thus satisfies~\eqref{eq:Bandedness}
in an ergodic sense, i.e., when replacing local averages $[ \,\cdots\, ]_{\mathrm{loc}}$ by ensemble averages $\av{ \,\cdots\, }$.

To arrive at a prediction for the perturbed dynamics $\langle A \rangle_{\!\rho_\lambda(t)}$, we first evaluate the average behavior $\av{ \langle A \rangle_{\!\rho_\lambda(t)} }$ over all members of the considered ensemble of perturbations.
Second, we consider the deviations $\xi_V(t) := \langle A \rangle_{\!\rho_\lambda(t)} - \av{ \langle A \rangle_{\!\rho_\lambda(t)} }$ for one particular realization from the average.
It turns out \cite{dab20relax} that the variance $\av{ \xi_V(t)^2 }$ is inversely proportional to the number $N_v$ of unperturbed levels mixed by the perturbation from assumption~\ref{lst:Assumptions:StrongPerturbation}.
Exploiting~\eqref{eq:NumMixedLevels}, we can therefore conclude that, for the overwhelming majority of individual perturbations in the considered ensemble, the actual behavior $\langle A \rangle_{\!\rho_\lambda(t)}$ is practically indistinguishable from the average $\av{ \langle A \rangle_{\!\rho_\lambda(t)} }$,
so that the latter in fact correctly describes the dynamics under nearly all perturbations of the ensemble for sufficiently large system sizes.
Results of this kind are also commonly known as ``typicality'', ``concentration of measure'', or ``ergodicity'' properties \cite{gog16, dal16, mor18}.

Taking for granted that the \sigmaName~\eqref{eq:Bandedness} is indeed 
the essential quantity for deviations between the perturbed and unperturbed systems, we may expect that also the behavior of the true system of interest should follow the ensemble average.
Unfortunately, it is hard to prove this for any given, concrete physical system.
Nevertheless, a phenomenological justification by means of examples is possible for a variety of different models \cite{dab20relax, dab20echo1}, see also Sec.~\ref{sec:Examples} below.
For the rest, we observe that the probability distribution~\eqref{eq:VProb} is still rather arbitrary since we only fix the first two moments of the densities $f_E(v)$.
In principle and if available, additional information about the distribution of the true $V_{\mu\nu}$ could thus be incorporated when choosing the $f_E(v)$,
but similarly as in the central limit theorem, these statistical properties turn out to be practically irrelevant,
reinforcing the pivotal role of the second moment~\eqref{eq:Bandedness}.

To conclude this section, we remark that the true perturbation will usually exhibit correlations (i.e., functional interdependencies) between the matrix elements $V_{\mu\nu}$, which may arise, for example, due to the locality and few-body character of interactions \cite{ham18, nic19}.
Since such correlations are not accounted for in the considered perturbation ensembles, it is implicitly assumed that their effect on the dynamics is negligible.
In practice, this particularly means that the reference Hamiltonian $H$ should be sufficiently ``clean'' such that the individual terms constituting the perturbation $V$ are in some sense ``orthogonal'' to those of $H$.
Notably, this rules out the possibility to ``reverse the roles'' by defining a new reference Hamiltonian $H' := H_\lambda$ and considering a perturbation $\lambda V' := H - H' = -\lambda V$ to predict $\<A\>_{\!\rho(t)}$ from $\< A \>_{\!\rho_\lambda(t)}$.

\section{Typical perturbed relaxation}
\label{sec:TypResponse}

Given the prominent role of the Hamiltonian as the generator of time evolution, it will be no surprise that the transformation matrices $U_{m\nu}$ between the eigenbases of $H$ and $H_\lambda$ [see Eq.~\eqref{eq:U}] are of particular importance to relate the unperturbed and perturbed dynamics.
Especially relevant turns out to be the so-called {\em \uName} $u(E)$, 
which describes the squared magnitude of the $U_{m\nu}$ 
averaged over the considered ensemble of perturbations,
\begin{equation}
\label{eq:u}
	\avv{ \lvert U_{m \nu} \rvert^2 } =: u(E_m - E_\nu) \,.
\end{equation}
Due to the (approximate) constancy of the level density 
[assumptions~\ref{lst:Assumptions:EnergyWindow} and~\ref{lst:Assumptions:WeakPerturbation}]
and the fact that the statistics of the $V_{\mu\nu}$ in (\ref{eq:VProb})
only depend on $E_{\mu}-E_{\nu}$, it follows that the
statistics of the $U_{m\nu}$ from~\eqref{eq:U} must be translationally invariant in energy, 
and hence the second moment in (\ref{eq:u}) must 
only depend on the energy difference
$E_m - E_\nu$.

Referring to Ref.~\cite{dab20relax} for the details, the typicality approach outlined
below Eq.~\eqref{eq:VProb} then eventually yields that, for the overwhelming majority of perturbations in any admissible ensemble~\eqref{eq:VProb}, the perturbed time evolution is given by
\begin{equation}
\label{eq:TypTimeEvo}
	\langle A \rangle_{\!\rho_\lambda(t)}
		= \langle A \rangle_{\!\tilde\rho} + \lvert g(t) \rvert^2 \left\{ \langle A \rangle_{\!\rho(t)} - \langle A \rangle_{\!\tilde\rho} \right\} .
\end{equation}
We recall that $\langle A \rangle_{\!\rho(t)}$ is the reference dynamics observed under 
the unperturbed Hamiltonian $H$.
Furthermore, the density operator $\tilde\rho$ appearing on the right-hand side of~(\ref{eq:TypTimeEvo}) is defined via its matrix elements $\bra{\mu} \tilde\rho \ket{\nu} := \delta_{\mu\nu} \sum_\kappa \tilde u(E_\nu - E_\kappa) \bra{\kappa} \rho(0) \ket{\kappa}$, where $\tilde u(E) := \int \d E' \, D(E') u(E - E') u(E')$.
In other words, $\tilde\rho$ may thus be viewed as the unperturbed diagonal ensemble associated with the initial state $\rho(0)$ which is in addition locally washed out via the function $\tilde u(E)$,
arising as the convolution of $u(E)$ with itself.
According to  \cite{gog16, deu91, rei15, nat18}, this operator $\tilde\rho$ can usually be well approximated by the microcanonical ensemble $\rhoMC$ corresponding to the pertinent energy window $I$ from \eqref{eq:DOS}.
Finally, the so-called {\em \gName} $g(t)$ on the right-hand side of (\ref{eq:TypTimeEvo}) is the Fourier transform of $u(E)$ from~\eqref{eq:u},
\begin{equation}
\label{eq:g}
	g(t) := \int \d E \, D(E) \, u(E) \, \e^{\I E t} \,.
\end{equation}
In particular, it can be readily verified that $\lvert g(0) \rvert^2 = 1$ and $\lvert g(t) \rvert^2 \to 0$ as $t \to \infty$.
According to~\eqref{eq:TypTimeEvo}, this function $g(t)$ thus describes how the unperturbed behavior is modified to approach the perturbed equilibrium value $\langle A \rangle_{\!\tilde\rho}$, i.e., it encodes the system's response to the perturbation.

The remaining task is to compute the function $u(E)$ from~\eqref{eq:u}.
To this end, we introduce the resolvent $\mathcal G(z) := (z - H_\lambda)^{-1}$ of $H_\lambda$,
which encodes the overlaps on the left-hand side of~\eqref{eq:u} as
$\lvert U_{m\nu} \rvert^2 \simeq \lim_{\eta\to 0+} \bra{\nu} [ \mathcal G(E_m - \I\eta) - \mathcal G(E_m + \I\eta) ] \ket{\nu} / 2\pi\I D(E_m)$ \cite{haa10, mir00}.
Since the ensemble average of $\mathcal G(z)$ can be written as $\av{ \mathcal G(z)  } = G(z - H)$ with the scalar function $G(z)$ defined in a minute, we can exploit $D(E_m) \approx \varepsilon^{-1}$ [cf.\ assumptions~\ref{lst:Assumptions:EnergyWindow} and~\ref{lst:Assumptions:WeakPerturbation}] to arrive at
\begin{equation}
\label{eq:uFromG}
	u(E) = \frac{\varepsilon}{\pi} \lim_{\eta\to 0+} \operatorname{Im} G(E - \I\eta) \,.
\end{equation}
Finally, the above introduced {\em ensemble-averaged resolvent} $G(z)$ itself can be obtained as the solution of the following nonlinear integral equation \cite{dab20relax, fyo96},
\begin{equation}
\label{eq:IntEqG}
	G(z) \left[ z - \lambda^2 \int \d E \, D(E) \, G(z - E) \, \varV(\lvert E \rvert) \right] = 1 \,.
\end{equation}

In summary, the strategy to obtain a prediction for the perturbed relaxation thus is to follow the sequence of equations~\eqref{eq:TypTimeEvo}--\eqref{eq:IntEqG} in reverse order:
First, for a given \sigmaName\ $\varV(E)$ and perturbation strength $\lambda$, we solve Eq.~\eqref{eq:IntEqG} for $G(z)$.
Second, this gives us access to the \uName\ $u(E)$ via Eq.~\eqref{eq:uFromG}.
Third, evaluating its Fourier transform~\eqref{eq:g} we obtain the
\gName\ $g(t)$, which then allows us, fourth, to predict $\langle A \rangle_{\!\rho_\lambda(t)}$ from the unperturbed $\langle A \rangle_{\!\rho(t)}$ according to Eq.~\eqref{eq:TypTimeEvo}.
Clearly, the first step, namely to solve the nonlinear integral equation \eqref{eq:IntEqG}, is the most demanding task. This problem is at the focus of the next section.

\section{Evaluation of the ensemble-averaged resolvent}
\label{sec:Resolvents}

In this section, we will discuss solutions $G(z)$ of Eq.~\eqref{eq:IntEqG} and the resulting \uName s
$u(E)$ from~\eqref{eq:uFromG}.
We first consider in Sec.~\ref{sec:Resolvents:Analytical} two limiting cases 
for which analytical approximations will be obtained.
Thereafter, we elaborate on how to solve Eq.~\eqref{eq:IntEqG} in the intermediate regime numerically using pseudospectral Chebyshev expansions \cite{for96, boy01}.
The evaluation of the predicted dynamics and its comparison with explicit examples is deferred to the ensuing Sec.~\ref{sec:Examples}.

\subsection{Analytically tractable special cases}
\label{sec:Resolvents:Analytical}

According to assumption~\ref{lst:Assumptions:Bandedness} from Sec.~\ref{sec:Scope},
the \sigmaName\ $\varV(E)$ 
from~\eqref{eq:Bandedness} is a well-behaving (continuous) function, so that the quantity
\begin{equation}
\label{eq:cvarV}
	\cmagV := \lim_{E\to 0+} \sqrt{ \varV(E) }
\end{equation}
exists \cite{note:diagV}. Essentially, $\cmagV$ thus characterizes
the ``intrinsic strength'' of the perturbations $V$.

As explained in Sec.~\ref{sec:Scope}, the perturbation matrix $V_{\mu\nu}$ in the 
eigenbasis of the unperturbed Hamiltonian $H$ is often expected to exhibit a banded 
structure, meaning that its \sigmaName\ $\varV(E)$ approaches zero as $E \to \infty$.
The corresponding so-called {\em ``band width''} or ``perturbation range'' may thus be quantified by
\begin{equation}
\label{eq:bwV}
	\bwV := \frac{1}{\cvarV} \int_0^\infty \d E \, \varV(E) \,.
\end{equation}
However, in full generality we will also admit cases where
$\varV(E)$ does {\em not} approach zero for large $E$.
In such a case, but also when $\varV(E)$ only decays very slowly 
with $E$, the band width $\bwV$ will be infinite.

Our first approximation starts from the observation that if the perturbation is 
sufficiently weak [sufficiently small $\lambda$ in (\ref{eq:H})] then also
the mixing of eigenvectors between the unperturbed and perturbed Hamiltonians 
should be weak in the sense that the concomitant eigenvector 
overlaps \eqref{eq:u} are only non-negligible for small energy differences 
$E_m-E_{\nu}$ of the corresponding eigenvalues.
In view of~\eqref{eq:uFromG}, we therefore
inspect the case
that the function $G(z-E)$ in the integrand in~\eqref{eq:IntEqG} exhibits (as a function of $E$, and for any preset $z$ of later relevance) a very narrow peak compared to
variations of the \sigmaName\ $\varV(E)$.
Accordingly, the integral is dominated by the region around the maximum of $G(z - E)$ at 
$E \approx \lvert z \rvert$, and we can approximate $\varV(\lvert E \rvert)$ by its central 
value $\varV(\lvert z \rvert)$.
Together with $D(E) \approx \varepsilon^{-1}$ [cf.\ assumption~\ref{lst:Assumptions:EnergyWindow}] we 
thus obtain
\begin{equation}
\label{eq:G:weak1}
	G(z) = \frac{1}{z - \lambda^2 \varV(\lvert z \rvert) C(z) / \varepsilon}
\end{equation}
with
\begin{equation}
\label{eq:G:weak1C}
	C(z) := \int \d E \, G(z - E) \,.
\end{equation}
Exploiting once again that $G(z)$ exhibits a very narrow peak compared to the variations of $\varV(|z|)$
implies with (\ref{eq:cvarV}) that $\varV(\lvert z \rvert) \approx \cvarV$ for all the relevant values 
of $\lvert z \rvert$ for which $G(z)$ significantly deviates from zero.
Furthermore, focusing in view of \eqref{eq:uFromG} on arguments $z$ of the form
$z=x-\I\eta$ with $x\in\RR$, the quantity $C(z)$ in (\ref{eq:G:weak1C}) assumes the same
constant value $C(-\I\eta)$ for all $z$.
In other words, $G(z)$ in (\ref{eq:G:weak1}) can be written as $1/(z-c)$ for some constant
$c\in\CC$.
Consequently, when evaluated in the principal value sense, $C(z)$ in~\eqref{eq:G:weak1C} only depends on the sign of the imaginary part of the denominator in~\eqref{eq:G:weak1}, yielding $C(z) = \mp \I \pi$ for $\operatorname{sgn}(\operatorname{Im} z) = \pm 1$ as the only consistent solution.
Altogether, we thus arrive at the approximation
\begin{eqnarray}
& & 	G(z) = \frac{1}{z + \I \operatorname{sgn}(\operatorname{Im} z) \, \Gamma/2 } \ ,
\label{eq:G:weak'}
\\
& & 	\Gamma :=  \frac{ 2\pi \lambda^2 \cvarV }{\varepsilon} \,,
\label{eq:G:weak}
\end{eqnarray}
and with~\eqref{eq:uFromG} we conclude that $u(E)$ approximately
assumes the Breit-Wigner form
\begin{equation}
\label{eq:u:weak}
	u(E) = \frac{\varepsilon}{2\pi} \frac{\Gamma}{E^2 + \Gamma^2 / 4} \,.
\end{equation}
Hence $\Gamma$ quantifies the peak width of $u(E)$,
and likewise for $G(z)$.
Our initial assumption that $G(z)$ is sharply peaked thus means that 
$\varV(E)$ must exhibit only small changes upon variations of
$E$ on the order of $\Gamma$.
Viewing the perturbation strength $\lambda$ as variable and all 
other system properties as fixed, we may thus consider
(\ref{eq:G:weak'})--(\ref{eq:u:weak}) as a weak perturbation 
approximation.
Importantly, this approximation is expected to apply for 
practically any reasonable \sigmaName\ $\varV(E)$
provided the perturbation strengths $\lambda$ are sufficiently small.
In many cases, one furthermore expects that the band width (\ref{eq:bwV})
at the same time quantifies the scale on which $\varV(E)$ exhibits notable
variations, yielding 
\begin{eqnarray}
\Gamma \ll \bwV
\label{c1}
\end{eqnarray} 
as the pertinent condition for the validity of the above approximations.
On the other hand, in cases where the variations of $\varV(E)$
remain relatively small for arbitrary $E$, those approximations
will actually apply to arbitrary coupling strengths $\lambda$
(apart from the general restrictions in Sec. \ref{sec:Scope}).

Our second approximation is similar in spirit but complementary to the first one.
Namely, we follow the same reasoning as before with the roles of $G(z)$ and 
$\varV(E)$ reversed, i.e., we now
consider the case
that the \sigmaName\ $\varV(E)$ 
is sharply peaked compared to the variations of $G(z)$
for arguments of the form $z=E-\I\eta$.
In the integrand in~\eqref{eq:IntEqG}, we thus approximate $G(z - E) \approx G(z)$ and (as before) $D(E) \approx \varepsilon^{-1}$, leading to
\begin{eqnarray}
& & 	
\gamma^2 G(z)^2 / 4 - z G(z) + 1 = 0 \ ,
\label{eq:G:strong1'}
\\
& & 
\gamma := \sqrt{8 \bwV / \varepsilon} \, \lambda \cmagV = \sqrt{ 4 \bwV \Gamma / \pi }
\ .
\label{eq:G:strong1}
\end{eqnarray}
Solving this algebraic equation for $G(z)$ and observing that $\operatorname{sgn}(\operatorname{Im} G(z)) = -\operatorname{sgn}(\operatorname{Im} z)$ due to $G(z) = \av{ (z - \lambda V)^{-1} }$ [see above~\eqref{eq:uFromG}], we obtain
\begin{equation}
\label{eq:G:strong}
	G(z) = \frac{2}{\gamma^2} \left[ z - \I \operatorname{sgn}(\operatorname{Im}z) \sqrt{ \gamma^2 - z^2 } \right] .
\end{equation}
Substituting into~\eqref{eq:uFromG}, we are left with the semicircular distribution
\begin{equation}
\label{eq:u:strong}
	u(E) = \frac{2 \varepsilon}{\pi\gamma^2} \sqrt{\gamma^2 - E^2} \, \Theta(\gamma^2 - E^2) \,,
\end{equation}
where $\Theta(x)$ denotes the Heaviside step function.
The condition
that $\varV(E)$ is sharply peaked thus means that 
$G(z=E-\I\eta)$ must exhibit only small changes upon variations of
$E$ on the order of $\gamma$.
Viewing the perturbation strength $\lambda$ as variable and all 
other system properties as fixed, we may thus consider
(\ref{eq:G:strong1})--(\ref{eq:u:strong}) as a strong perturbation 
approximation.
More precisely, this approximation is expected to apply for 
practically any reasonable \sigmaName\ $\varV(E)$
provided the perturbation strengths $\lambda$ are 
sufficiently large, and provided that $\varV(E)$ does 
approach zero for large $E$ in the first place.
In particular, this is the case if the band width $\bwV$
from (\ref{eq:bwV}) is finite.
Furthermore, if $\bwV$ at the same time quantifies the scale on which 
$\varV(E)$ exhibits notable variations, 
then the pertinent condition for
the validity of the above approximations
assumes the form 
\begin{eqnarray}
\gamma \gg \bwV
\ .
\label{c2}
\end{eqnarray} 

Essentially, the \uName\ $u(E)$ from~\eqref{eq:u} 
is thus predicted to approximately assume the Breit-Wigner form~\eqref{eq:u:weak} under the weak perturbation condition (\ref{c1}), and the semicircular form~\eqref{eq:u:strong} under the strong perturbation condition
(\ref{c2}), largely independently of any further details of the \sigmaName\ $\varV(E)$ from~\eqref{eq:Bandedness}.
In the intermediate regime, characterized by $\Gamma \simeq \gamma$, or equivalently
\begin{equation}
\label{eq:lambdaCrit}
	\lambda \simeq \lambda_{\mathrm{c}} := \frac{ \sqrt{2 \varepsilon \bwV } }{\pi \cmagV} \,,
\end{equation}
one thus expects a smooth crossover between these limiting cases (see also Fig.~\ref{fig:u} below),
which will depend on the detailed shape of the \sigmaName\ $\varV(E)$,
and which in general will only be tractable by numerical means.

\begin{figure*}
\includegraphics[scale=1]{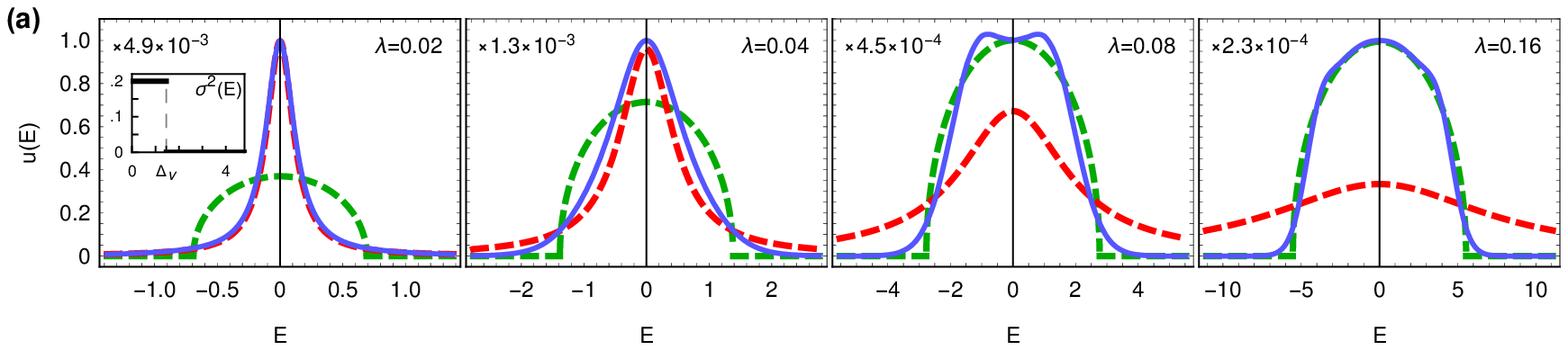}\\
\vspace{5pt}
\includegraphics[scale=1]{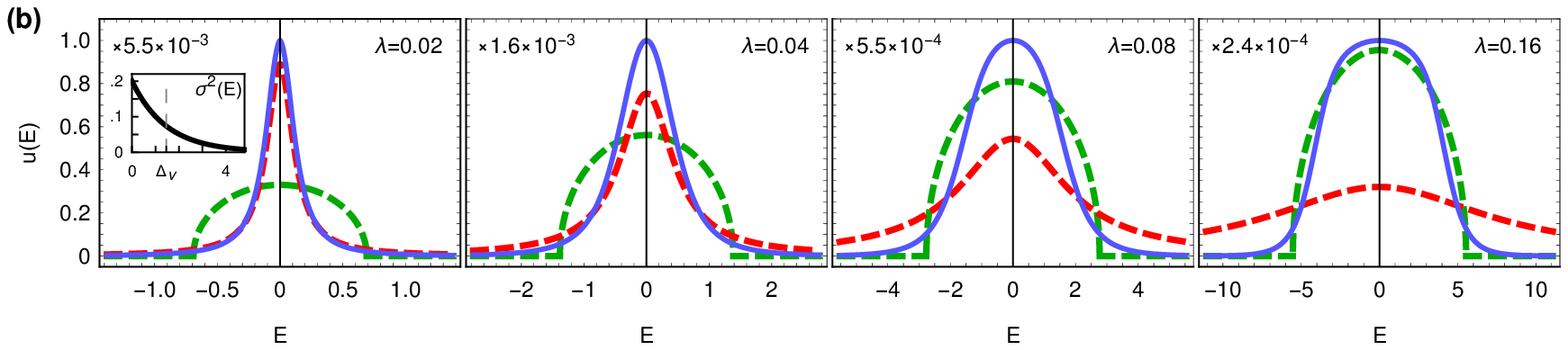}\\
\vspace{5pt}
\includegraphics[scale=1]{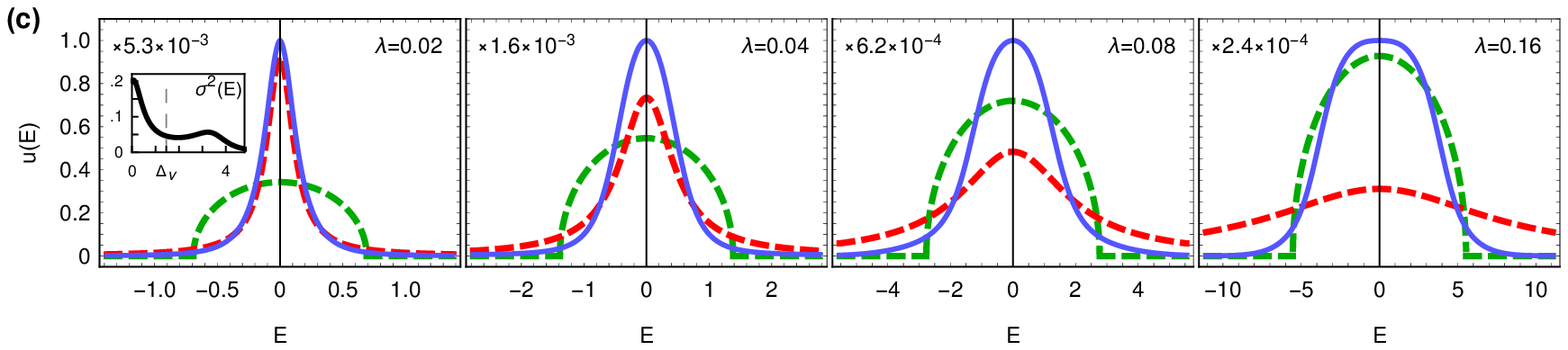}\\
\caption{Overlap distribution $u(E)$ from~\eqref{eq:u} 
for three different \sigmaName s $\varV(E)$ as depicted in the insets of the left-most panel of each row, namely
a step profile~\eqref{eq:BandProfile:Step} in (a), an exponential profile~\eqref{eq:BandProfile:Exp} in (b), 
and a double-Breit-Wigner profile~\eqref{eq:BandProfiles:DoubleBW} 
with $b_1 = 0.45$, $b_2 = 0.9$, $d = 3.5$ in (c).
In all three cases, we employed the same parameter values
$\varepsilon = 1/512$, 
$\cvarV = 0.2$, and
$\bwV  = 750\lvsp = 1.46$.
In each row, the perturbation strength $\lambda$ is increased from left to right as specified in the top-right corner of each panel.
Solid blue lines correspond to the numerical solution of 
\eqref{eq:uFromG} via~\eqref{Geta}--\eqref{eq:ChebGExpansion},
while dashed lines show the limiting Breit-Wigner (red/dark) and semicircular (green/light) distributions according to Eqs.~\eqref{eq:u:weak} and~\eqref{eq:u:strong}, expected for weak and strong perturbations, respectively.
As predicted below~\eqref{eq:lambdaCrit}, the crossover between these two limits occurs around $\lambda \simeq \lambda_{\mathrm{c}} \approx 0.05$.
Note that the vertical axes are scaled as indicated in the top-left corner of each panel.
}
\label{fig:u}
\end{figure*}

\subsection{Numerical treatment of the general case}
\label{sec:Resolvents:Numerical}

Our goal is to determine the \uName\ $u(E)$ according to \eqref{eq:uFromG}
by numerically solving the nonlinear integral equation \eqref{eq:IntEqG} for
largely general \sigmaName s $\varV(E)$.
As in the previous subsection,
we thus can and will focus in \eqref{eq:IntEqG} on arguments $z$ of the form
$z=x-\I\eta$ with $x\in\RR$ and $\eta > 0$ very small.
As noted below~\eqref{eq:G:strong1}, the relation $G(x-\I\eta) = \av{(x - \I\eta -\lambda V)^{-1}}$ implies $\operatorname{Im} G(x-\I\eta) \geq 0$ for $\eta > 0$ and vice versa, i.e., the sign of the imaginary part of $G(z)$ jumps when crossing the real line.
For purely real $z$, in turn, this implies that the solution of~\eqref{eq:IntEqG} becomes ambiguous, 
depending on whether one chooses to continue from the upper or lower half-plane.
Bearing in mind that the latter option is appropriate in \eqref{eq:uFromG},
we introduce the abbreviation
\begin{eqnarray}
G_+(x) := \lim_{\eta\to 0+} G(x-\I\eta)
\ .
\label{Geta}
\end{eqnarray}
Exploiting (as usual) that
$D(E) \approx \varepsilon^{-1}$ 
[cf.\ assumption~\ref{lst:Assumptions:EnergyWindow}],
the integral equation \eqref{eq:IntEqG} can thus be rewritten 
for real-valued $x$ as
\begin{equation}
\label{eq:IntEqG'}
	G_{+}(x) \left[ x - \frac{\lambda^2}{\varepsilon} \int \d E \, G_{+}(x - E) \, \varV(\lvert E \rvert) \right] = 1 
\end{equation}
with the additional constraint that
\begin{equation}
\label{eq:GPlus:Constraint}
	\operatorname{Im} G_+(x) \geq 0 \,.
\end{equation}

Our method of choice to solve Eq.~\eqref{eq:IntEqG'} numerically is an expansion in terms of Chebyshev rational functions $B_n(x)$ ($n = 0, 1, \ldots$),
which are derived from the Chebyshev polynomials of the first kind $T_n(x)$ by a compactification of the real line,
\begin{equation}
\label{eq:ChebRationalFunctions}
	B_n(x) := T_n\!\left( \frac{x}{\sqrt{x^2 + \ell^2}} \right) .
\end{equation}
Here $\ell$ is an arbitrary, fixed parameter that sets the scale for compactification and should roughly reflect the typical scale of the function to be expanded for optimal convergence \cite{boy01}.
Hence we express
\begin{equation}
\label{eq:ChebGDecomposed}
	G_{+}(x) = G^{\mathrm R}(x) + \I G^{\mathrm I}(x) \,,
\end{equation}
where the real-valued functions $G^{\mathrm R}(x)$ and $G^{\mathrm I}(x)$ are truncated Chebyshev series, i.e.,
\begin{equation}
\label{eq:ChebGExpansion}
	G^{\mathrm R}(x) := \sum_{n=0}^M G^{\mathrm R}_n B_n(x)
	\;\; \text{ and } \;\;
	G^{\mathrm I}(x) := \sum_{n=0}^M G^{\mathrm I}_n B_n(x) \,.
\end{equation}
The (real-valued) coefficients $G^{\mathrm R}_n$ and $G^{\mathrm I}_n$ are then to be determined such that 
$G_{+}(x)$ from~\eqref{eq:ChebGDecomposed} satisfies~\eqref{eq:IntEqG'} and \eqref{eq:GPlus:Constraint}
``as well as possible.''
For given expansion coefficients $\bm G := (G^{\mathrm R}_0, G^{\mathrm I}_0, \ldots, G^{\mathrm R}_M, G^{\mathrm I}_M)$ and $x$, the residual [i.e., the violation of Eq.~\eqref{eq:IntEqG}] is defined as
\begin{equation}
\label{eq:ChebResidual}
	R(\bm G, x) := G_{+}(x) \left[ x - \frac{\lambda^2}{\varepsilon} \int \d E\, G_{+}(x - E) \varV(\lvert E \rvert) \right] - 1
\end{equation}
with $G_{+}(x)$ from~\eqref{eq:ChebGDecomposed} and~\eqref{eq:ChebGExpansion}.
We minimize $\lvert R(\bm G, x) \rvert^2$ by means of pseudospectral methods \cite{for96, boy01}, requiring $\operatorname{Re} R(\bm G, x_m) = \operatorname{Im} R(\bm G, x_m) = 0$ for a discrete set of 
real-valued collocation points $x_m$ ($m = 0, 1, \ldots, M$).
A common choice for these $x_m$ is to use the roots of the $(M+1)$th Chebyshev rational function $B_{M+1}(x)$, so that the pseudospectral method coincides with a spectral expansion when an optimal Gaussian quadrature rule is used to calculate inner products numerically \cite{boy01, for96}.

Altogether, forcing $\operatorname{Re} R(\bm G, x_m) = \operatorname{Im} R(\bm G, x_m) = 0$ results in a set of $2(M+1)$ algebraic equations for the $2(M+1)$ unknown expansion coefficients 
$G^{\mathrm R}_n,\, G^{\mathrm I}_n \in\RR$.
This system of equations is then solved iteratively by the Newton-Raphson method using either of the limiting distributions~\eqref{eq:u:weak} or~\eqref{eq:u:strong} for the first initial guess, and gradually varying $\lambda$ across the intermediate regime thereafter.
If the initial guess is sufficiently close to the actual solution and 
satisfies $\operatorname{Im} G_+(x) \geq 0$, this ensures that also the 
finally obtained approximation will fulfill the constraint~\eqref{eq:GPlus:Constraint}.

In Fig.~\ref{fig:u}, we display the so-obtained 
numerical solutions $u(E)$ in~\eqref{eq:uFromG} for different \sigmaName s $\varV(E)$ and various perturbations strengths $\lambda$ along with the limiting Breit-Wigner 
functions~\eqref{eq:u:weak} expected for small $\lambda$ and 
the semicircular functions~\eqref{eq:u:strong} expected for large $\lambda$.
The selected \sigmaName s are a step function,
\begin{equation}
\label{eq:BandProfile:Step}
	\varV(E) = \cvarV \, \Theta( \bwV - E ) \,,
\end{equation}
an exponential function,
\begin{equation}
\label{eq:BandProfile:Exp}
	\varV(E) = \cvarV \, \e^{-E / \bwV} \,,
\end{equation}
and a double-Breit-Wigner function,
\begin{equation}
\label{eq:BandProfiles:DoubleBW}
	\varV(E) = \cvarV \frac{ b_1^2 (b_2^2 + d^2) }{ (b_1^2 + E^2) [ (b_2^2 + (E - d)^2 ] } \,.
\end{equation}
All three \sigmaName s are also shown in the insets of the left panels in Fig.~\ref{fig:u}.
Parameters are chosen such that in all cases
$\varepsilon = 1/512$ (mean level spacing), 
$\cvarV = 0.2$ (cf.~Eq.~(\ref{eq:cvarV})),
and $\bwV  = 750\lvsp = 1.46$ (band width, cf.~Eq.~\eqref{eq:bwV}), 
yielding a value of $\lambda_{\mathrm c} \approx 0.05$ for the 
crossover coupling strength in (\ref{eq:lambdaCrit}).
Moreover, the order of the Chebyshev expansions is $M = 80$ throughout, with the parameter $\ell$ varying between $0.5$ and $8$ [roughly optimizing the global residual~\eqref{eq:ChebResidual}].

For each of the three profiles~\eqref{eq:BandProfile:Step}--\eqref{eq:BandProfiles:DoubleBW},
the predicted crossover from the Breit-Wigner to the semicircular shape of $u(E)$
is clearly visible as $\lambda$ is increased.
The intermediate regime, where neither the Breit-Wigner nor the semicircular distribution offers a satisfactory approximation, appears to be somewhat smaller for the discontinuous step profile than for the smooth exponential and double-Breit-Wigner profiles.
In any case, in this intermediate regime there is a (relatively mild) dependence of $u(E)$ on the detailed shape of $\varV(E)$.
It therefore seems reasonable to expect that -- at least in principle -- it may be possible
to reconstruct from a sufficiently precisely known function $u(E)$ the 
underlying \sigmaName\ $\varV(E)$.


\section{Evaluation of the relaxation dynamics and examples}
\label{sec:Examples}

With our above obtained results for the \uName\ $u(E)$ at hand, we now turn to their implications for the \gName\ 
$g(t)$, which governs the deviations of the perturbed from the unperturbed 
relaxation behavior according to~\eqref{eq:TypTimeEvo}.
Specifically, we will first address in Sec.~\ref{sec:Examples:g} some more general issues, 
while in the subsequent Secs.~\ref{sec:Examples:RandomMatrix} and~\ref{sec:Examples:Spin}, 
we will compare our theoretical prediction~\eqref{eq:TypTimeEvo} with 
two explicit examples of random-matrix and spin models, respectively.


\begin{figure*}
\includegraphics[scale=1]{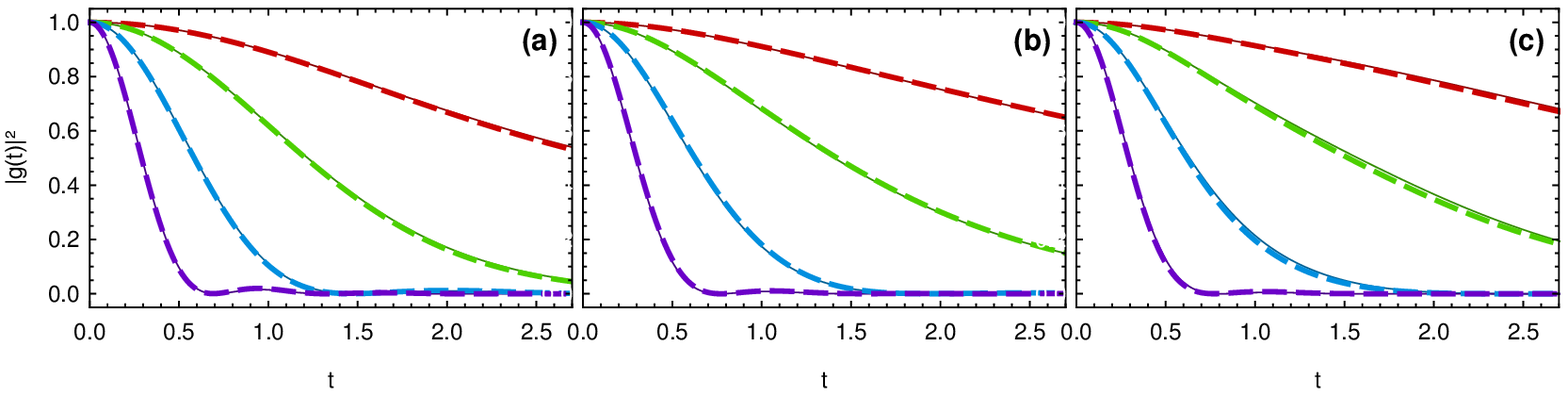}\\
\caption{Theoretical prediction (solid) versus random-matrix simulation (dashed) of the function 
$|g(t)|^2$
from~\eqref{eq:TypTimeEvo} and~\eqref{eq:g} 
for the same examples as in Fig.~\ref{fig:u},
namely a step \sigmaName~\eqref{eq:BandProfile:Step} in (a),  
an exponential profile~\eqref{eq:BandProfile:Exp} in (b), 
and a double-Breit-Wigner profile~\eqref{eq:BandProfiles:DoubleBW} 
with $b_1 = 0.45$, $b_2 = 0.9$, $d = 3.5$ in (c),
and 
$\varepsilon = 1/512$, 
$\cvarV = 0.2$,
$\bwV  = 750\lvsp = 1.46$.
The values for lambda are $\lambda = 0.02, 0.04, 0.08, 0.16$ in each panel (similar to Fig.~\ref{fig:u}), increasing from top to bottom.
}
\label{fig:g}
\end{figure*}


\subsection{Response profile}
\label{sec:Examples:g}

Exploiting in ~\eqref{eq:g} our usual approximation
$D(E) \approx \varepsilon^{-1}$ [cf.\ assumption~\ref{lst:Assumptions:EnergyWindow}],
the \gName\ $g(t)$ can be readily obtained via Fourier transformation 
from our analytical and numerical findings for $u(E)$ in the 
previous Sec.~\ref{sec:Resolvents}.
For the two analytically tractable special cases from Sec.~\ref{sec:Resolvents:Analytical}, 
the Fourier transformation can again be performed analytically, 
whereas for the numerical solutions from Sec.~\ref{sec:Resolvents:Numerical}, 
also the Fourier transformation is only possible by numerical means.

In the limit of weak perturbations, 
when $u(E)$ assumes the Breit-Wigner form~\eqref{eq:u:weak}, 
one readily finds along these lines that
$g(t)$ amounts to an exponential decay,
\begin{equation}
\label{eq:g:weak}
	g(t) = \e^{-\Gamma \lvert t \rvert / 2} \,,
\end{equation}
where the rate $\Gamma$ is the full width at half maximum of $u(E)$ as defined in~\eqref{eq:G:weak}.

Likewise, for (moderately) strong perturbations such that $u(E)$ takes 
the semicircular shape~\eqref{eq:u:strong}, its Fourier transform is
\begin{equation}
\label{eq:g:strong}
	g(t) = \frac{2 J_1(\gamma t)}{\gamma t} \,,
\end{equation}
where $J_1(x)$ is the Bessel function of the first kind of order $1$, and 
$\gamma$
as specified in Eq.~\eqref{eq:G:strong1} is the radius of the semicircle.

In the intermediate regime, our findings for $u(E)$ imply that $g(t)$ must exhibit
a crossover between these two limiting behaviors.
Calculating the Fourier transforms of the numerical solutions for $u(E)$ from Fig.~\ref{fig:u}, 
we obtain the solid curves shown in Fig.~\ref{fig:g} for $\lvert g(t) \rvert^2$,
which is the actually relevant quantity in~\eqref{eq:TypTimeEvo}.
This illustrates quantitatively the expected
crossover from~\eqref{eq:g:weak} to~\eqref{eq:g:strong} 
with increasing $\lambda$.

The first general conclusion is that the perturbed relaxation becomes 
faster with increasing $\lambda$.
Quite obviously, the underlying physical reason is
a corresponding broadening of $u(E)$ with increasing $\lambda$,
which in turn indicates (as expected) that an increased number 
of unperturbed energy levels are coupled by the perturbation
according to~(\ref{eq:u}).

The second general conclusion is that the functions $g(t)$
become independent of any further details of the \sigmaName\ $\varV(E)$ for asymptotically large or small $\lambda$,
while some (rather moderate) functional dependence on 
$\varV(E)$ remains in the intermediate regime.
Again, the underlying reasons are our analogous observations
for the \uName s $u(E)$ in the preceding section.
Though the functional dependence of $g(t)$, and thus of the 
perturbed relaxation in~\eqref{eq:TypTimeEvo},
is quantitatively rather weak, it still may be possible, at least in principle,
to infer the (coarse-grained) \sigmaName~\eqref{eq:Bandedness} 
of the specific perturbation $V$ for some given many-body system
(\ref{eq:H}) from the observable temporal relaxation 
via~\eqref{eq:TypTimeEvo}.

\subsection{Random matrix example}
\label{sec:Examples:RandomMatrix}

To verify that the theoretical prediction~\eqref{eq:TypTimeEvo} indeed describes the behavior of many-body quantum systems (provided that assumptions~\ref{lst:Assumptions:EnergyWindow} 
through~\ref{lst:Assumptions:Bandedness}
from Sec.~\ref{sec:Scope} hold),
we finally compare it to explicit numerical examples.

The first example is a (in some sense artificial) random matrix model that satisfies the requirements from Sec.~\ref{sec:Scope} by construction and thus serves as a testbed for the validity of the approximations employed in the derivation of Eq.~\eqref{eq:TypTimeEvo} (see also Ref.~\cite{dab20relax}).
The reference Hamiltonian has equally spaced energy levels $E_\nu = \nu \varepsilon$ with $\varepsilon = 1/512$.
The perturbation $V$ is a complex Hermitian random matrix distributed according to~\eqref{eq:VProb} with
\begin{equation}
\label{eq:RandMat:VProb}	
	f_E(v) = (1 - p) \, \delta(v) + \frac{p \, \e^{ -\lvert v \rvert^2 / \hat\sigma^2(E)} }{\pi \hat\sigma^2(E)} 
	\quad (E > 0) \,.
\end{equation}
On average, the matrices $V_{\mu\nu}$ are thus sparse with a fraction $p$ of nonvanishing entries following a complex normal distribution of variance $\hat\sigma^2(E)$ for $\mu < \nu$, and $V_{\nu\mu} = V^*_{\mu\nu}$.
For simplicity, the diagonal matrix elements $V_{\nu\nu}$ are sampled similarly, but with a real normal distribution for the nonvanishing entries.
Consequently, the \sigmaName~\eqref{eq:Bandedness} is given by
\begin{equation}
\label{eq:RandMat:Bandedness}
	\varV(E) = p \, \hat\sigma^2(E) \,.
\end{equation}
Specifically, we implemented the three 
\sigmaName s~\eqref{eq:BandProfile:Step}--\eqref{eq:BandProfiles:DoubleBW} 
with $\cvarV = p = 0.2$ and $\bwV = 1.46$ (corresponding to about $750$ levels).

The initial state $\rho(0) = \ket{\nu_0} \bra{\nu_0}$ is an eigenstate of the reference 
Hamiltonian $H$ from the middle of the spectrum, and we observe its survival 
probability or fidelity \cite{gor06, tor14quench}, i.e., $A = \rho(0)$.
Hence $\langle A \rangle_{\!\rho(t)} = 1$ for all $t$ while $\langle A \rangle_{\!\tilde\rho} = \Ath \approx 0$ for a sufficiently large energy window $I$
from \eqref{eq:DOS}, so that the prediction~\eqref{eq:TypTimeEvo} reduces to
\begin{equation}
\label{eq:RandMat:TypTimeEvo}
	\langle A \rangle_{\!\rho_\lambda(t)} = \lvert g(t) \rvert^2 \,.
\end{equation}
In other words, recording the dynamics in this setup for one particular perturbation sampled from~\eqref{eq:RandMat:VProb}, we should exactly recover the solid curves in Fig.~\ref{fig:g}.
The dashed lines in the figure represent one such example dynamics for a 
Hilbert space of dimension $2^{14} = 16\,384$ 
and an initial eigenstate $\ket{\nu_0}$ with $\nu_0 = 2^{13} = 8192$.

The main conclusion is that the simulation results indeed 
agree almost perfectly with the theoretically predicted 
solid curves throughout the entire crossover regime.

\subsection{Spin lattice example}
\label{sec:Examples:Spin}

\begin{figure*}
\includegraphics[scale=1]{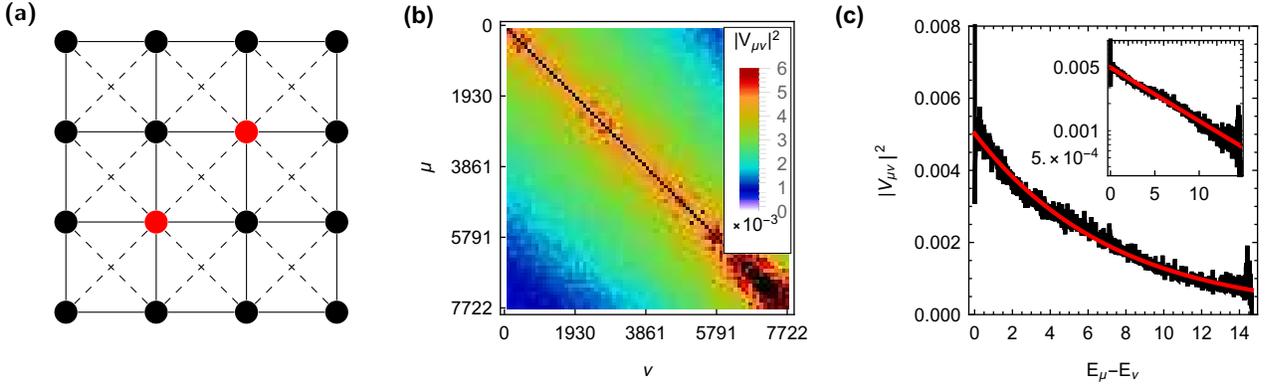}
\caption{
(a) Illustration of the spin model~\eqref{eq:Spin:H}--\eqref{eq:Spin:V}.
Solid links correspond to sites coupled via the reference Hamiltonian $H$, 
dashed links to those coupled by the perturbation $V$.
Highlighted in red are the sites $(2,2)$ and $(3,3)$, on which the two considered observables 
[see Eqs.~\eqref{eq:Spin:obsMagCorr} and~\eqref{eq:Spin:obsHoppingCorr}]
are supported.
(b) Squared matrix elements $\lvert V_{\mu\nu} \rvert^2$ of the perturbation~\eqref{eq:Spin:V} in the eigenbasis of the reference Hamiltonian~\eqref{eq:Spin:H} in a central energy window $I$ of $7722$ states ($60\,\%$ of the total) in the zero-magnetization sector, averaged over blocks of $100\times 100$ levels.
(c) Coarse-grained \sigmaName~\eqref{eq:Bandedness} (black, bin width $0.01$) and fit to the exponential form~\eqref{eq:BandProfile:Exp} with $\cvarV = 0.00502$ and $\bwV = 7.32$ (red).
The inset shows the same data with a logarithmically scaled $y$-axis.}
\label{fig:SpinModel}
\end{figure*}

\begin{figure*}
\includegraphics[scale=1]{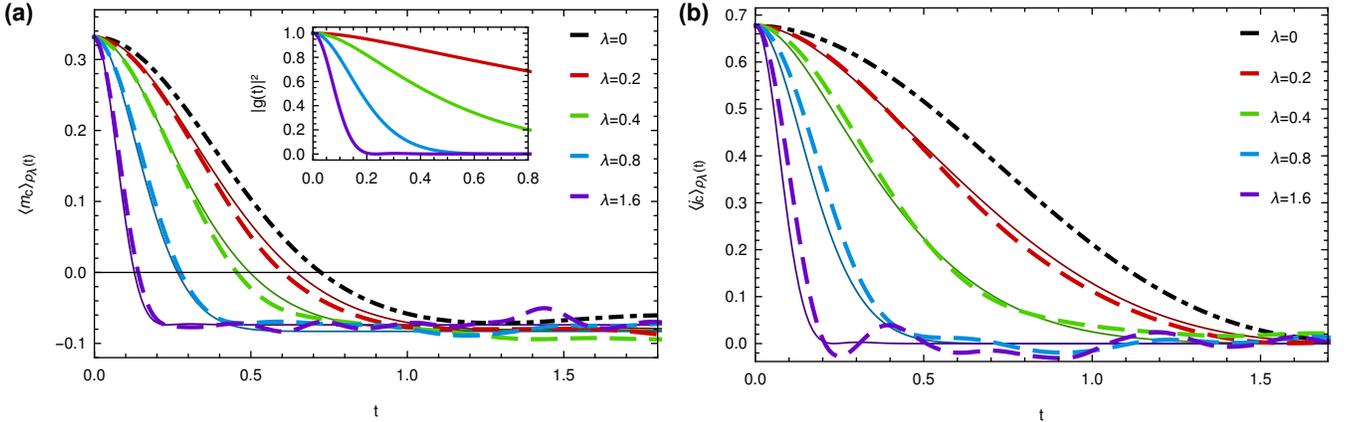}\\
\caption{Time-dependent expectation values of (a) the central magnetization 
correlation $m_{\mathrm c}$ from~\eqref{eq:Spin:obsMagCorr} and (b) the central 
spin-flip correlation $j_{\mathrm c}$ from~\eqref{eq:Spin:obsHoppingCorr} for the 
spin system with $H_\lambda = H + \lambda V$ from~\eqref{eq:Spin:H}--\eqref{eq:Spin:V} 
and various perturbation strengths $\lambda$ as indicated, increasing from top to bottom. 
The initial state $\rho(0) = \ket{\psi}\bra{\psi}$ is 
chosen according to~\eqref{eq:Spin:initStateMagCorr} 
in (a) and~\eqref{eq:Spin:initState} in (b).
Dashed lines represent the numerical values obtained by exact diagonalization.
Solid lines correspond to the theoretical prediction from~\eqref{eq:TypTimeEvo} using the (black, dash-dotted) $\lambda = 0$ curve as input for the reference dynamics $\langle A \rangle_{\!\rho(t)}$. The \gName\ $g(t)$ (see inset of left panel) is calculated according to Secs.~\ref{sec:TypResponse} and~\ref{sec:Resolvents} from the \sigmaName\ $\varV(E)$ from Fig.~\ref{fig:SpinModel}(c).
For the long-time average $\langle A \rangle_{\!\tilde\rho}$, the explicit prediction for the state $\tilde\rho$ [see below Eq.~\eqref{eq:TypTimeEvo}] is used in (a), yielding $\langle A \rangle_{\!\tilde\rho} = -0.0896, -0.0820, -0.0830, -0.0738$ for $\lambda = 0.2, 0.4, 0.8, 1.6$, respectively.
In (b), $\tilde\rho$ is taken to be thermal, $\langle A \rangle_{\!\tilde\rho} = \Ath = 0$.
}
\label{fig:SpinDynamics}
\end{figure*}

Finally, we test the theoretical prediction~\eqref{eq:TypTimeEvo} in a more realistic two-dimensional spin-$\frac{1}{2}$ model.
We consider a square lattice of $L \times L$ sites as sketched in Fig.~\ref{fig:SpinModel}(a), where the reference Hamiltonian couples nearest neighbors with an isotropic spin-spin interaction,
\begin{equation}
\label{eq:Spin:H}
	H = J \sum_{i,j=1}^{L-1} \bm\sigma_{i,j} \cdot \left( \bm\sigma_{i+1,j} + \bm\sigma_{i,j+1} \right) .
\end{equation}
Here $\bm\sigma_{i,j} := (\sigma^x_{i,j}, \sigma^y_{i,j}, \sigma^z_{i,j})$ with $\sigma^\alpha_{i,j}$ denoting the Pauli matrices acting on site $(i,j)$.
The perturbation adds 
spin-flip terms between next-nearest neighbors,
\begin{equation}
\label{eq:Spin:V}
	V = \sum_{i,j=1}^{L-1} \sum_{\alpha = x,y} \left( \sigma^\alpha_{i,j} \sigma^\alpha_{i+1,j+1} + \sigma^\alpha_{i+1,j} \sigma^\alpha_{i,j+1} \right) .
\end{equation}
In all of the numerics presented here, we used $L = 4$ and $J = 1$, and 
we focused on the sector with vanishing total magnetization 
in the
$z$-direction.

To obtain the \sigmaName~\eqref{eq:Bandedness} of $V$, we first fix an energy 
window $I$ by choosing the central $60\,\%$ of energy levels, which comprises a total of $7722$ states ranging from $E = -8.8$ to $E = 5.8$, implying a mean level spacing $\varepsilon = 0.0019$.
Next we compute the matrix elements $V_{\mu\nu}$ with $E_\mu, E_\nu \in I$ by diagonalizing the reference Hamiltonian $H$.
A coarse-grained view of the resulting matrix is shown in Fig.~\ref{fig:SpinModel}(b), visualizing the bandedness of the perturbation matrix.
We proceed by binning the $V_{\mu\nu}$ according to the energy difference $E_\mu - E_\nu$ of the associated levels and evaluate the average of $\lvert V_{\mu\nu} \rvert^2$ within each bin.
The obtained relation between the coarse-grained $\lvert V_{\mu\nu} \rvert^2$ and $E_\mu - E_\nu$ is displayed as a black curve in Fig.~\ref{fig:SpinModel}(c), indicating an approximately exponential dependency.
The function $\varV(E)$ is then determined by fitting the exponential form~\eqref{eq:BandProfile:Exp} to the empirical distribution, yielding the red line in Fig.~\ref{fig:SpinModel}(c) with $\cvarV = 0.00502$ and $\bwV = 7.32$.
This implies a value of $\lambda_{\mathrm c} = 0.75$ for the predicted location of the crossover~\eqref{eq:lambdaCrit} between the exponential and Bessel-type decay characteristics~\eqref{eq:g:weak} and~\eqref{eq:g:strong}, respectively.

As a first observable, we investigate the magnetization correlation $m_{\mathrm c}$ 
in the $z$ direction between next-nearest neighbors from the center of the lattice,
\begin{equation}
\label{eq:Spin:obsMagCorr}
	m_{\mathrm c} := \sigma^z_{2,2} \sigma^z_{3,3} \,.
\end{equation}
One could consider these two spins at $(2,2)$ and $(3,3)$ as the system 
and all other surrounding spins as a bath.
In the reference Hamiltonian $H$, the system spins can thus only interact 
via the bath, whereas the perturbation $V$ adds a direct interaction between them.

For the initial state $\rho(0) = \ket{\psi} \bra{\psi}$, we choose 
those two system spins at $(2,2)$ and $(3,3)$ to be
in the ``up'' state, while the bath is supposed to be at equilibrium, 
which we emulate by choosing a Haar-distributed random vector in the bath's subspace.
However, to ensure assumption~\ref{lst:Assumptions:EnergyWindow} of a well-defined macroscopic energy, we finally apply a Gaussian projection $\Pi_{\mathcal{E}, \Delta_{\mathcal{E}}}$ of mean energy $\mathcal{E} = 0$ and standard deviation $\Delta_{\mathcal{E}} = 2$ to the so-obtained state, simulating a macroscopic measurement of the system energy that yielded $\mathcal E = 0$ \cite{pre95, gar13, ste14}.
If $\ket{\phi}$ denotes a Haar-distributed random vector on the full (zero-magnetization) Hilbert space, we thus have
\begin{equation}
\label{eq:Spin:initStateMagCorr}
	\ket{\psi} \propto \Pi_{\mathcal{E}, \Delta_{\mathcal{E}}} \sigma^+_{2,2} \sigma^{+}_{3,3} \ket{\phi} 
\end{equation}
with $\sigma^+_{i,j} := \sigma^x_{i,j} + \I \sigma^y_{i,j}$ and
\begin{equation}
\label{eq:Spin:GaussFilter}
	\Pi_{\mathcal{E}, \Delta_{\mathcal{E}}} \propto \sum_\nu \e^{-(E_\nu - \mathcal{E})^2 / 2 \Delta_{\mathcal{E}}^2 } \ket{\nu} \bra{\nu} \,.
\end{equation}

In Fig.~\ref{fig:SpinDynamics}(a), we compare the observed dynamics obtained by exact diagonalization (dashed lines) with our theoretical prediction~\eqref{eq:TypTimeEvo} (solid lines) for several perturbation strengths $\lambda$.
For the theoretical prediction, we use the numerical reference dynamics (i.e., the dash-dotted black curve with $\lambda = 0$) for $\langle m_{\mathrm c} \rangle_{\!\rho(t)}$.
The function $g(t)$ is the Fourier transform of $u(E)$ calculated as explained in Sec.~\ref{sec:Resolvents:Numerical} from the empirically determined approximate \sigmaName\ $\varV(E)$, i.e., the red curve in Fig.~\ref{fig:SpinModel}(c).
The so-obtained \gName s $g(t)$ are also displayed in the inset of Fig.~\ref{fig:SpinDynamics}(a).
Since the long-time limiting values exhibit some finite-size variations, we do not use the microcanonical value $\langle m_{\mathrm c} \rangle_{\!\mathrm{mc}} = -0.0805$ (within the $60\,\%$ window $I$) 
for $\langle m_{\mathrm c} \rangle_{\!\tilde\rho}$,
but instead compute the predicted coarse-grained diagonal ensemble $\tilde\rho$ directly as detailed below Eq.~\eqref{eq:TypTimeEvo}, making use of our solution for $u(E)$ and the known occupations $\bra{\nu} \rho(0) \ket{\nu}$ of the initial state from~\eqref{eq:Spin:initStateMagCorr}.
The resulting quantitative values of $\langle m_{\mathrm c} \rangle_{\!\tilde\rho}$
for the various perturbations strengths $\lambda$ are given in the figure caption.

The agreement between theory and numerics is very good despite the rather small system size and several idealizations.
In particular, the assumptions of a homogeneous density of states [assumption~\ref{lst:Assumptions:EnergyWindow}], of an exponential perturbation profile [assumption~\ref{lst:Assumptions:Bandedness} and Fig.~\ref{fig:SpinModel}(c)], 
and of uncorrelated matrix elements $V_{\mu\nu}$
[see above (\ref{eq:VProb})]
are all violated to some extent and are thus potential origins of the visible small deviations in Fig.~\ref{fig:SpinDynamics} for short times.
The fluctuations for longer times, in contrast, are likely caused predominantly by finite-size effects.
We
emphasize that there are no free parameters in the theoretical prediction;
all ingredients in~\eqref{eq:TypTimeEvo} were extracted directly from properties of the model~\eqref{eq:Spin:H}--\eqref{eq:Spin:V}.

As a second observable, we consider the spin-flip or hopping correlation $j_{\mathrm c}$ 
between the same sites $(2,2)$ and $(3,3)$ from the center of the lattice 
in Fig.~\ref{fig:SpinModel}(a),
\begin{equation}
\label{eq:Spin:obsHoppingCorr}
\begin{aligned}
	j_{\mathrm c} :=& \, \sigma^x_{2,2} \sigma^y_{3,3} - \sigma^y_{2,2} \sigma^x_{3,3} \\
		=& \, \frac{1}{2\I} \left( \sigma^-_{2,2} \sigma^+_{3,3} - \sigma^+_{2,2} \sigma^-_{3,3} \right) ,
\end{aligned}
\end{equation}
where $\sigma^{\pm}_{i,j} := \sigma^x_{i,j} \pm \I \sigma^y_{i,j}$.
For the initial state, we employ a dynamical typicality setup \cite{bar09a, rei18} 
to prepare the system far from equilibrium, choosing
\begin{equation}
\label{eq:Spin:initState}
	\ket{\psi} \propto \Pi (1 + \kappa j_{\mathrm c}) \Pi \ket{\phi} \,,
\end{equation}
where $\ket{\phi}$ is a Haar-distributed random state as before, $\Pi$ is a 
projector onto the central $2048$ states in the zero-magnetization sector [ensuring assumption~\ref{lst:Assumptions:EnergyWindow}], and $\kappa$ is a real 
parameter (in the examples, we use $\kappa = 2$).

A similar comparison as for $m_{\mathrm c}$ between numerical simulations and the theoretical prediction~\eqref{eq:TypTimeEvo} is shown for the hopping correlation $j_{\mathrm c}$ from~\eqref{eq:Spin:obsHoppingCorr} in Fig.~\ref{fig:SpinDynamics}(b).
In particular, the functions $g(t)$ are the same in both panels of Fig.~\ref{fig:SpinDynamics}.
On the other hand, in this setup $\langle j_{\mathrm c} \rangle_{\!\tilde\rho}$ is well approximated by the thermal expectation value $\langle j_{\mathrm c} \rangle_{\!\mathrm{mc}} = 0$ (by symmetry), so that we used this value throughout.
Altogether, this amounts again to an entirely parameter-free prediction of the perturbed dynamics, which agrees well with the actually observed behavior.

\section{Conclusions}
\label{sec:Conclusions}

We investigated the response of quantum many-body systems to weak-to-moderate perturbations within a nonperturbative typicality framework.
In particular, we presented a method to theoretically predict time-dependent expectation values of 
observables for the perturbed system from the unperturbed relaxation behavior.
This prediction~\eqref{eq:TypTimeEvo} entails that the perturbed relaxation resembles the unperturbed one, but is modified by a characteristic response profile function $g(t)$ that pushes the system towards a coarse-grained diagonal ensemble state, which can usually be identified with the pertinent thermal state.
The function $g(t)$, in turn, is essentially determined by the \sigmaName, i.e., the locally averaged squared absolute value~\eqref{eq:Bandedness} of the perturbation's matrix elements $V_{\mu\nu}$ in the unperturbed basis.

For asymptotically weak perturbations, the \gName\ $g(t)$ describes 
an exponential decay, where the decay rate corresponds to the energy scale across 
which the perturbation mixes unperturbed eigenstates,
scaling quadratically with 
the perturbation strength $\lambda$.
Broadly speaking, this may be understood as a nonperturbative justification of 
Fermi's golden rule in a many-body setting.

The nonperturbative character of our method becomes manifest as the perturbation 
strength is increased.
Our results then predict a crossover of $g(t)$ towards the Bessel-type 
shape~\eqref{eq:g:strong}, whose 
inverse relaxation time scale
$\gamma$ still quantifies 
the mixing of energy levels, but now scales linearly with $\lambda$ and additionally
depends on the energy range $\bwV$ of the perturbation.

We verified all those theoretical predictions in an explicit example of a spin system on a $4 \times 4$ square lattice.
Using exact diagonalization to determine the \sigmaName\ of the applied perturbation empirically [cf.\ Fig.~\ref{fig:SpinModel}(c)], the function $g(t)$ derived from it indeed describes the actually observed perturbed dynamics remarkably well as long as the key assumptions~\ref{lst:Assumptions:EnergyWindow} 
through~\ref{lst:Assumptions:Bandedness}
collected in Sec.~\ref{sec:Scope} are satisfied.
Notably, the theory does not involve any free parameters, i.e., all quantities were determined first-hand from the underlying spin model.
Since the \sigmaName\ is the only variable input for the theory, this establishes that said profile
encodes the dynamical response on a fundamental level.

Then again, the correspondence between the \sigmaName\ and the dynamical response may in principle be exploited the other way round, too.
The rapidly improving experimental capabilities to observe time-dependent expectation values of mesoscopic quantum systems may thus offer a way to probe the (coarse-grained) matrix elements of applied perturbations.
A similar proposal to extract matrix structures from dynamics can also be found in the recent work \cite{mal19heating} using periodic driving and working in the regime of weak perturbations governed by the exponential law~\eqref{eq:g:weak}.
Our present approach can be considered complementary in that it avoids time-dependent manipulations and extends to significantly stronger perturbations.
Given the important role of matrix elements in the energy eigenbasis for the dynamics in general and for questions of equilibration and thermalization (e.g.\ the eigenstate thermalization hypothesis) in particular, this sets up new possibilities to explore the underlying mechanisms by means of time series analysis.

\begin{acknowledgments}
This work was supported by the 
Deutsche Forschungsgemeinschaft (DFG)
within the Research Unit FOR 2692
under Grants No.~397303734
and No.~397300368
and by the Paderborn Center for Parallel 
Computing (PC$^2$) within the Project 
HPC-PRF-UBI2.
\end{acknowledgments}


\begin{thebibliography}{70}
\bibitem{note:GroundState}
We note that initial states close to the ground state (or the upper end of the spectrum, if it exists) are implicitly excluded due to assumption~\ref{lst:Assumptions:EnergyWindow},
so that the potentially missing confinement at the edges of the spectrum (due to less pronounced level repulsion) does not impede the argument.
\bibitem{note:StateBroadening}
In view of the sentence around Eq.~\eqref{eq:U}, the initial state may typically populate 
a wider range of levels in the perturbed system than in the unperturbed one, i.e., the width 
of the distribution of occupied levels may increase by an amount on the order of $\Gamma$.
As detailed in \cite{dab20relax}, this implies that the width $\Delta$ 
of the energy window $I$ introduced above Eq.~\eqref{eq:DOS} must be 
sufficiently large and $\Gamma$ must remain sufficiently small
so that $\Gamma \ll \Delta$.
\bibitem{f1}
We note that this assumption is somewhat similar to (but still considerably weaker than)
the (off-diagonal) eigenstate thermalization hypothesis (ETH) 
\cite{deu91,sre94,rig08,dal16}.
\bibitem{note:diagV}
The variance of the diagonal elements $V_{\nu\nu}$ need not be of the same order as that of the close-by off-diagonal elements, i.e., $\varV(E)$ from~\eqref{eq:Bandedness} need not be continuous as $E \to 0$ \cite{fyo95}.
\bibitem{f2}
On the other hand, the maximum of the function $\varV(E)$ 
in (\ref{eq:Bandedness}) is 
tacitly assumed not to significantly exceed its values in the vicinity 
of $E=0$, again in agreement with all specific examples we know of.
\end{thebibliography}
\end{document}